\documentclass[intlimits,twoside,a4paper]{article}
\usepackage{graphicx}
\usepackage{wrapfig}
\usepackage[T2A]{fontenc}
\usepackage[cp1251]{inputenc}

\usepackage{cmpj2}

\issue{2014}{17}{3}{33802}
\doinumber{10.5488/CMP.17.33802}
\title
{Transmission of cultural traits in layered ego-centric networks}%

\author{V. Palchykov\refaddr{label1,label2,label3},
        K. Kaski\refaddr{label1,label4,label5}, J. Kert\'esz\refaddr{label6,label1}}

\addresses{
\addr{label1} Department of Biomedical Engineering and Computational Science, Aalto University, \\00076 Aalto, Finland
\addr{label2} Institute for Condensed Matter Physics, National Academy of Sciences of Ukraine, 79011 Lviv, Ukraine
\addr{label3} Lorentz Institute for Theoretical Physics, Leiden University, 2300 RA Leiden, The Netherlands
\addr{label4} CABDyN Complexity Center, Said Business School, University of Oxford, Oxford OX1 1HP, UK
\addr{label5} CCNR and Physics Department, Northeastern University, Boston, MA 02115, USA
\addr{label6} Center for Network Science, Central European University, Nador 9, H--1051, Budapest, Hungary
}

\date{Received May 22, 2014, in final form June 12, 2014}
\begin{document}

\maketitle

\begin{abstract}
Although a number of models have been developed to investigate the emergence
of culture and evolutionary phases in social systems, one important aspect
has not yet been sufficiently emphasized. This is the structure of the underlaying network of social relations serving as
channels in transmitting cultural traits, which is expected to play a crucial role in the evolutionary processes in social systems. In this paper we contribute to the understanding of the role of the network structure by developing a layered ego-centric network structure based model, inspired by the social brain hypothesis, to study transmission of cultural traits and their evolution in social network. For this model we first find analytical results in the spirit of mean-field approximation and then to validate the results we compare them with the results of extensive numerical simulations.

\pacs 89.75.Fb, 89.65.Ef, 89.75.Hc
\end{abstract}

\section{Introduction}\label{sect_I}
Emergent natural phenomena like ferromagnetism in materials, and social phenomena like formation of collective opinion or evolution of culture in a society have clearly different origins. However, during the last two decades it has become increasingly clear that there are important common features, namely they are all complex systems, in which a large number of interacting units contribute to entirely new cooperative qualities. Formulating this way, it is not anymore surprising that concepts, methods, and modelling techniques of statistical physics have increasingly been applied to a  number of problems beyond the traditional realm of physics, giving rise to such new discipline names as ``econophysics'' \cite{Mantegna2000} and ``sociophysics'' \cite{Sen2013,Castellano2009}.

In physics sense, modelling means conceptual simplification, in which one aims to find a small number of assumptions to capture the essence of the phenomenon investigated. This is exemplified by a model of two possible spin states for each particle, like the Ising model for magnet, which in spite of its simplicity is capable of describing spontaneous symmetry breaking and the related emergence of magnetization. Similarly, simple models of disease spreading with two possible states (either ``susceptible'' or ``infected'') for the members of the society permit to describe the emergence of a pandemic state in the system. A major difference between modelling natural and social phenomena is, however, that in the former case general principles like variational laws or microscopic equations of motion are usually known, while in the latter case of social systems, the situation is more complicated. In this case, one uses plausible assumptions about the processes leading to stochastic rules, which are then tested by comparing the results with reality - usually in a qualitative manner.

Here we will discuss, from the sociophysics perspective, the problem of competition and coexistence of two cultural traits in a society. The situation is in a sense similar to binary alloys, in which for some parameters the components mix, while for the others a phase separation takes place. This analogy raises the question about a kind of cultural phase diagram, which is a known problem in cultural anthropology \cite{Axelrod1997,Centola2007}.  While much of research has been done from the mean field point of view, only recently the topology of the interactions has been taken into account \cite{Guerra2010}.

The natural way to describe the processes of transmission of values in a social system is to assume transition rates for the evolution of the local properties by which Markov chain equations can be formulated for the behavior of the entire system. While in this context the Markovian approximation has its obvious limitations, it is still worth studying for several reasons. First, the Markovian approximation should hold for a large time scale; second, the solutions can serve as references and third, this is the technique we are most familiar with.

In our present study we focus on modelling a biased transmission of cultural traits,
developed to describe the evolution of culture in a society \cite{Boyd1985}.
Among different types of biased transmission there is direct bias with one
cultural trait being more attractive than the others or indirect bias with an
individual using a characteristic not immediately related to the cultural trait
(e.g., individual success). Here we focus on the former one, namely on direct bias.
The assumption that one cultural trait is more attractive than the other,
will cause its frequency to increase in time. However, the effect of this
behavior depends on the network of social interactions that allow the cultural
traits to be transmitted between the members of a social system. Here we adopt
the view of ego-centric social networks to describe the emotional closeness of the
social brain hypothesis \cite{Dunbar1998}, which besides the restriction on the number
of active social network members for each individual (the Dunbar number of about 150) suggests the entire ego-centric network consisting of the layers that correspond to
different levels of intensities of interactions or emotional closeness \cite{Dunbar1995,Kudo2001,Zhou2005}.

In this paper we first describe the biased transmission model in section~\ref{sect_II}; develop a mean-field approximation for the evolution of a cultural trait in the entire system with the given ego-centric network in section~\ref{sect_III}; construct two distinct cases of entire networks with a fixed ego-centric structure in section~\ref{sect_IV}; and investigate the applicability of the derived mean field approximation. In section~\ref{sect_V} we draw conclusions.

\section{Transmission models with direct bias}\label{sect_II}

Assuming that there are two possible cultural traits $x_i(t) = \{ 0, 1\}$ for each individual $i=1,\ldots, N$ in a system, transmission models allow each individual to change his or her cultural trait in time $t$ as a result of adopting it from available sources. The model of direct bias suggests some bias in choosing and adopting cultural trait, which means that one of the traits is more attractive. Then we introduce a quenched quantity $b$ to determine the strength of this bias, the value of which varies within the range $b\in (-1,1)$. For $b>0$, individuals are predisposed towards choosing trait $1$ while for $b<0$ cultural trait $0$ is favoured. The possible sources of cultural trait for each ego $i$ correspond to the ``friends'' of individual $i$ ego-centric social network and to the ego $i$ him- or herself.

Let us now assume that the ego-centric social network of individual $i$ consists of $N_i$ friends and that the strength of relation between the ego $i$ and source $j$ is $\omega_{ij}$. It reflects the influence of each of the source $j$ on the ego $i$. The values of $\omega_{ij}$ are normalized
\begin{equation}\label{eq2.1}
\sum_{\langle j\rangle_i}\omega_{ij} = 1,
\end{equation}
where $\langle{j}\rangle_i$ means that $j$ runs over the sources of the cultural trait for ego $i$.

For the direct bias transmission model \cite{Boyd1985}, each ego $i$ decides to choose one of his or her source of cultural trait and adopts the trait from that source. The decision which source to choose depends both on $\omega_{ij}$ and on the combination of the bias $b$ and the cultural trait $x_j(t)$ of the source $j$. The probability $q_{ij}$ that the ego $i$ adopts cultural trait from the source $j$ is determined \cite{Boyd1985} as
\begin{equation}\label{eq2.2}
q_{ij} = \frac{\omega_{ij}(1+\beta_j)}{\sum_{\langle{k}\rangle_i} \omega_{ik}(1+\beta_k)}\,,
\end{equation}
where
\begin{equation}\label{eq2.3}
\beta_j= \left\{\begin{array}{rrl}
  b, & {\rm for} & x_j(t)=1,\\
 -b, & {\rm for} & x_j(t)=0.\\
  \end{array}\right.
\end{equation}
In equation~(\ref{eq2.2}) $k$ runs over all available sources of a cultural trait for ego $i$ (including the ego him- or herself, i.e., $k=i$). Note that the value of $q_{ij}$ in equation~(\ref{eq2.2}) depends on the whole configuration $\{x_j(t)\}_i$ of cultural traits $x_j(t)$ of all the sources for ego $i$.
Then the probability $p_{i}(t+1|\{x_j(t)\}_i)$ that an individual $i$ will adopt cultural trait $x_i(t+1) = 1$ at time $t+1$ given the configuration $\{x_j(t)\}_i$ at time $t$ may be expressed as follows:
\begin{equation}\label{eq2.4}
 p_{i}(t+1|\{x_j(t)\}_i) = \sum_{\langle{j}\rangle_i} q_{ij}x_j(t).
\end{equation}
Equation (\ref{eq2.4}) provides a general rule for the evolution of individual cultural trait. The fraction of individuals $p(t)$, as characterized by cultural trait 1 at time $t$ for the large enough population follows
\begin{equation}\label{eq2.5}
 p(t) = \frac{1}{N}\sum_{i=1}^N p_{i}(t),
\end{equation}
where $p_i(t)$ is the probability that an individual $i$ has cultural trait $x_i(t) = 1$ at time $t$.
To investigate the evolution of a cultural trait in the system one has to specify both the structure of the network of social interactions and the initial conditions. Below we consider two types of ego-centric social networks and apply a mean-field approximation that allows us to analytically estimate the evolution of a cultural trait in the entire system.

\section{Mean-field approximation}\label{sect_III}

Here, we investigate two cases of social networks that reflect possible sources of a cultural trait. First, we consider a simple ego-centric network \cite{Boyd1985}, where each individual has only one external source of cultural trait at each time $t$, but this source changes over time and may be considered as a randomly chosen individual in a system. Second, we consider layered ego-centric network, assuming that the number of sources within each layer is large enough. The mean field approximations are obtained assuming that the probability $p_i(t)$ of having a cultural trait $x_i(t)=1$ is independent of $i$, and thus $p_i(t) = p(t)$. Below we consider the preference towards cultural trait 1, i.e., $b>0$.

\subsection{Single external source of a cultural trait}

Let us assume that each individual $i$ has only two possible sources of a cultural trait labelled with $j$: (i) himself or herself ($j=i$) or (ii) an external source ($j\neq i$), so that the strength $\omega_{ij}$ of social influence of the source $j$ reads as follows:
\begin{equation}\label{eq3.1}
\omega_{ij}= \left\{\begin{array}{ll}
  \alpha_0,	& \mbox{if $j=i$,}\\
  1-\alpha_0,	& \mbox{if $j\neq i$.}\\
  \end{array}\right.
\end{equation}
In order to reproduce a temporal evolution of an individual cultural trait [equation~(\ref{eq2.4})] the entire configuration $\{x_j(t)\}_i$ of cultural traits $x_j(t)$ of all the sources for ego $i$ must be specified.
In the approximation used, the exact configuration $\{x_j(t)\}_i$ is not available and we deal with the probability $P(\{x_j(t)\}_i)$ of realization of each particular configuration $\{x_j(t)\}_i$. Then equation~(\ref{eq2.4}) may be expanded as follows:
\begin{equation}\label{eq3.2}
p_i(t+1) = \sum_{\{x_j(t)\}_i}\Big[P(\{x_j(t)\}_i)\sum_{\langle{j}\rangle_i} q_{ij}x_j(t)\Big],
\end{equation}
where the first sum runs over four possible configurations of $\{x_j(t)\}_i$.
Recalling that $x_j(t)$ for each source $j$ of cultural trait at time $t$ is as follows:
\begin{equation}\label{eq3.3}
x_j(t)= \left\{\begin{array}{ll}
  1, & \mbox{with probability $p(t)$,}\\
  0, & \mbox{with probability $1-p(t)$,}\\
  \end{array}\right.
\end{equation}
one may simply calculate the values of $P(\{x_j(t)\}_i)$ for each of the four possible configurations $\{x_j(t)\}_i$. Substituting these values into equation~(\ref{eq3.2}) and taking the strength of social influence of each source $j$ on ego $i$, $\omega_{ij}$, into account [equation~(\ref{eq3.1})], one obtains an estimation for the evolution of individual cultural trait $p_i(t+1)$, which in the mean-field approximation is equal to the fraction of population $p(t+1)$ [equation~(\ref{eq2.5})] characterised by a favoured cultural trait \cite{Boyd1985}:
\begin{equation}\label{eq3.4}
 p(t+1) = p(t)+p(t)\big[1-p(t)\big]\frac{4b\alpha_0(1-\alpha_0)}{1-b^2(2\alpha_0-1)^2}\,.
\end{equation}

The result (\ref{eq3.4}) was obtained in 1980-s and it shows that the system starting from arbitrary small number of individuals with cultural trait $1$ reaches a final state in which all individuals are characterized by the favored trait $1$ for an arbitrarily small value of bias $b>0$. A more complicated structure of a social network (decision among a number of sources) makes the calculations harder, and numerical simulations were then used to investigate the system behavior. R.I.M.~Dunbar \cite{Dunbar2011} considered the behavior of the system where each ego has two layers of sources of a cultural trait, and where the strength of relations decays with the layer number. Numerical calculations demonstrated a qualitatively similar but sufficiently faster spreading of the cultural trait in a system when more sources are simultaneously considered by the ego.

\subsection{Layered ego-centric network}

Here we develop a mean-field approximation for the evolution of a cultural trait in a system with a layered ego-centric social network. Let us assume that there are two layers of friends that correspond to different levels of emotional closeness: Let each individual have $n_1$ first- and $n_2$ second layer friends. Then the strength of social influence can be written as follows:
\begin{equation}\label{eq3.5}
\omega_{ij}= \left\{\begin{array}{ll}
  \alpha_0, & \mbox{if $i=j$},\\
  \alpha_1, & \mbox{if $j$ is the first-layer friend for ego $i$},\\
  \alpha_2, & \mbox{if $j$ is the second-layer friend for ego $i$},\\
  \end{array}\right.
\end{equation}
and satisfy the normalization condition (\ref{eq2.1}). If the number of sources within each layer is large enough, then $n_lp(t)$ and $n_l[1-p(t)]$ approximate the numbers of $l$-th layer friends with the cultural trait equal to 1 and 0, respectively. Assuming that $n_lp(t)$ and $n_l[1-p(t)]$ match the corresponding numbers, the influence of the first- and second-layer friends on the ego $i$ may be considered as an effective eternal field; and this will allow us to avoid considering all possible configurations $\{x_j(t)\}_i$ of cultural traits, as it is shown below. Within this assumption, $q_{ij}$ among the whole configuration $\{x_j(t)\}_i$ of cultural traits depends only on two of them, namely on $x_i(t)$ and $x_j(t)$ being implemented into $\beta_i$ and $\beta_j$, correspondingly:
\begin{equation}\label{eq3.6}
q_{ij} = \frac{\omega_{ij}(1+\beta_j)}{\alpha_0(1+\beta_i) + \sum_{l=1}^2\{\alpha_ln_lp(t)(1+b) + \alpha_ln_l[1-p(t)](1-b)\}}\,.
\end{equation}
The normalization condition (\ref{eq2.1}) being substituted into equation~(\ref{eq3.6}) allows us to exclude both $\alpha_1n_1$ and $\alpha_2n_2$ from the equation (\ref{eq3.6}):
\begin{equation}\label{eq3.7}
q_{ij} = \frac{\omega_{ij}(1+\beta_j)}{\alpha_0(1+\beta_i) + [1-\alpha_0]\{p(t)(1+b) + [1-p(t)](1-b)\}}\,.
\end{equation}

To investigate the temporal evolution of a cultural trait, equation (\ref{eq3.2}) is used again. Within the approximation used here, the nested sum of this equation becomes dependent only on the cultural trait $x_i(t)$ of ego $i$:
\begin{equation}\label{eq3.8}
\sum_{\langle{j}\rangle_{i}} q_{ij} x_j(t) = \frac{\alpha_0(1+\beta_i)x_i(t) + [1-\alpha_0]p(t)(1+b)}{\alpha_0(1+\beta_i) + [1-\alpha_0]\{p(t)(1+b) +[1-p(t)](1-b)\}}\,.
\end{equation}
Now to estimate $p_i(t+1)$, it is enough to consider only two possible values of $x_i(t)$ among the whole set of configurations $\{x_j(t)\}_i$ in equation~(\ref{eq3.2}).
If $x_i(t)=1$, then the probability that an individual $i$ conserves the cultural trait at time $t+1$ follows
\begin{equation}\label{eq3.9}
\sum_{\langle{j}\rangle_i} q_{ij}{x}_j(t)\Big|_{x_i(t)=1}= \frac{\alpha_0(1+b) + [1-\alpha_0]p(t)(1+b)}{\alpha_0(1+b) + [1-\alpha_0]p(t)(1+b) + [1-\alpha_0][1-p(t)](1-b)}\,.
\end{equation}
Instead, if $x_i(t)=0$ the probability that the individual $i$ will adopt a cultural trait $x_i(t+1)=1$ at time $t+1$ follows
\begin{equation}\label{eq3.10}
\sum_{\langle{j}\rangle_i} q_{ij}{x}_j(t)\Big|_{x_i(t)=0} = \frac{ [1-\alpha_0]p(t)(1+b)}{\alpha_0(1-b) + [1-\alpha_0]p(t)(1+b) + [1-\alpha_0][1-p(t)](1-b)}\,.
\end{equation}

According to equation~(\ref{eq3.2}), the probability $p_i(t+1)$ that an individual $i$ is characterized by a cultural trait $1$ at time $t+1$ as follows:
\begin{equation}\label{eq013}
 p_i(t+1) = p(t) \sum_{\langle{j}\rangle_i} q_{ij}{x}_j(t)\Big|_{x_i(t)=1} + [1-p(t)] \sum_{\langle{j}\rangle_i} q_{ij}{x}_j(t)\Big|_{x_i(t)=0}.
\end{equation}
Substituting equations~(\ref{eq3.9}) and (\ref{eq3.10}) into equation~(\ref{eq013}) and taking into account that $p_i(t)=p(t)$, one obtains the mean-field approximation for the fraction of the population with the cultural trait favored by bias as follows:
\begin{equation}\label{eq014}
 p(t+1) = p(t)+p(t)(1-p(t))(1-\alpha_0)\Big[\frac{1+b}{\Sigma(t)-\alpha_0b} - \frac{1-b}{\Sigma(t)+\alpha_0b}\Big],
\end{equation}
where
\begin{equation}
 \Sigma(t) = 1 - (1-\alpha_0)b[1-2p(t)].
\end{equation}

Under the assumption that the numbers $n_1$ and $n_2$ of first- and second layer friends are large enough to apply the mean-field approximation and due to the normalization condition (\ref{eq2.1}), the result (\ref{eq014}) obtained does not depend either on $\alpha_1$ or on $\alpha_2$. Only the value of $\alpha_0$ is important. For similar reasons, the existence of additional layers does not change the result (\ref{eq014}) provided that the number of sources within each layer is large enough.
To verify the validity of the mean-field approximation (\ref{eq014}),  in the next section we compare it with numerical simulations.

\section{Numerical simulations}\label{sect_IV}

Finally, we perform numerical simulations of the biased transmission model on layered ego-centric networks in order to test the results of the mean-field approximation (\ref{eq014}). According to the social brain hypothesis \cite{Dunbar1998}, the capacity of a human brain limits the number of individuals with whom one can maintain real social relationships. After reaching that limit of about 150 acquaintances, the quality of relationships changes significantly. Naturally, within this set of acquaintances there are large differences as regards emotional closeness of the acquaintances to an ego. R.A.~Hill and R.I.M.~Dunbar (see \cite{Hill2003} and references therein) state that different layers of emotional closeness exist, creating a set of inclusive layers, the sizes of which increase in factors of about 3 with each layer, starting from a support clique of $\sim 5$, to a sympathy group of $\sim 12$, to a band of $\sim 35$ and to a cognitive group of $\sim 150$) individuals. Moreover, these layers of emotional closeness are correlated with the frequency of contacts between individuals \cite{Roberts2010}, so that an ego contacts with a member of a support clique on a daily basis, with a member of a sympathy group on a weekly basis, with a member of a band on monthly basis and with a member of a cognitive group on a yearly basis. For our numerical simulations we will implement this ego-centric network structure to investigate the transmission of a cultural trait in social networks, as we describe herein below.

\subsection{Network generation}
\begin{figure}[!ht]
\includegraphics[width=0.48\textwidth]{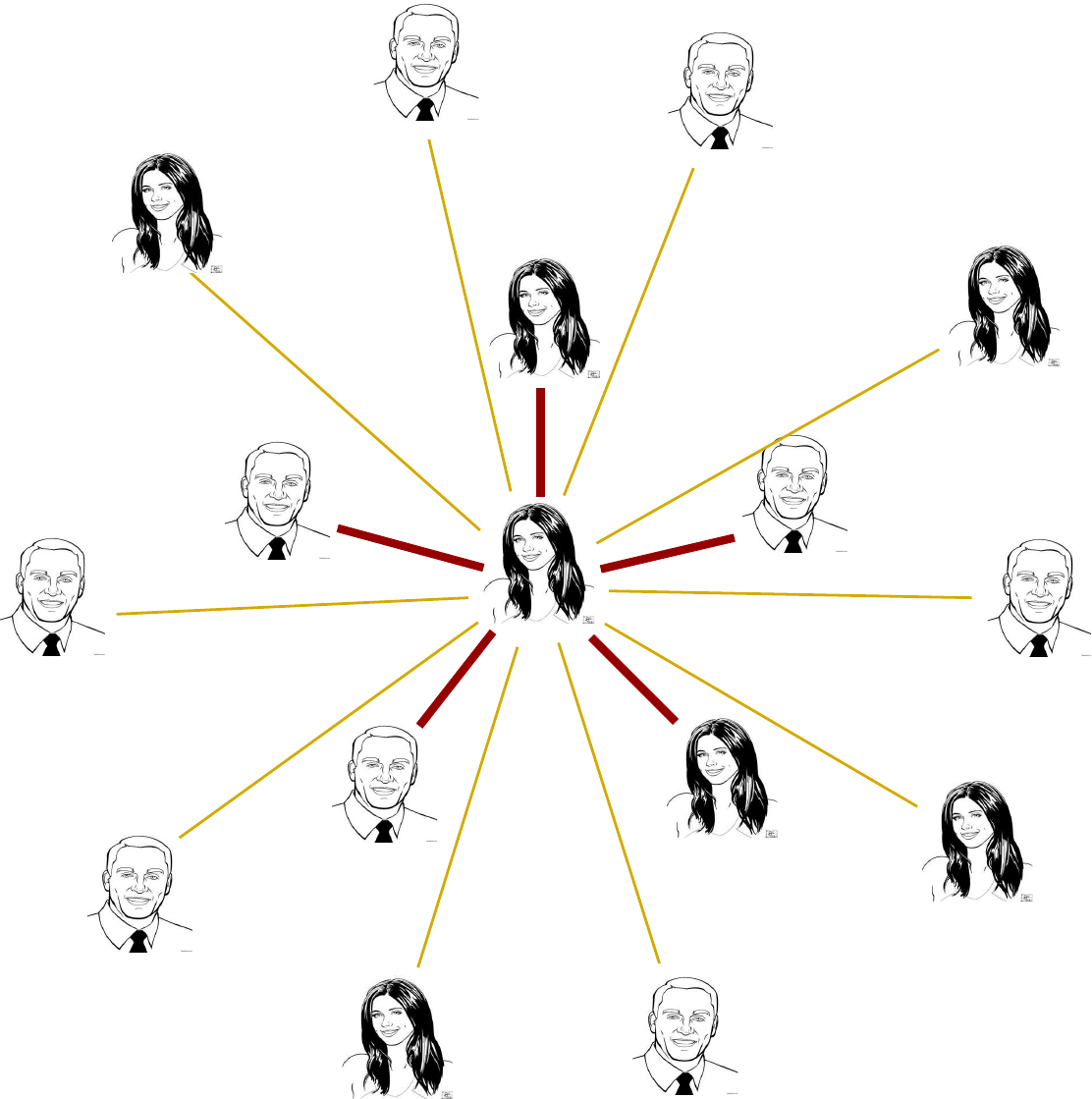}%
\hfill%
\includegraphics[width=0.48\textwidth]{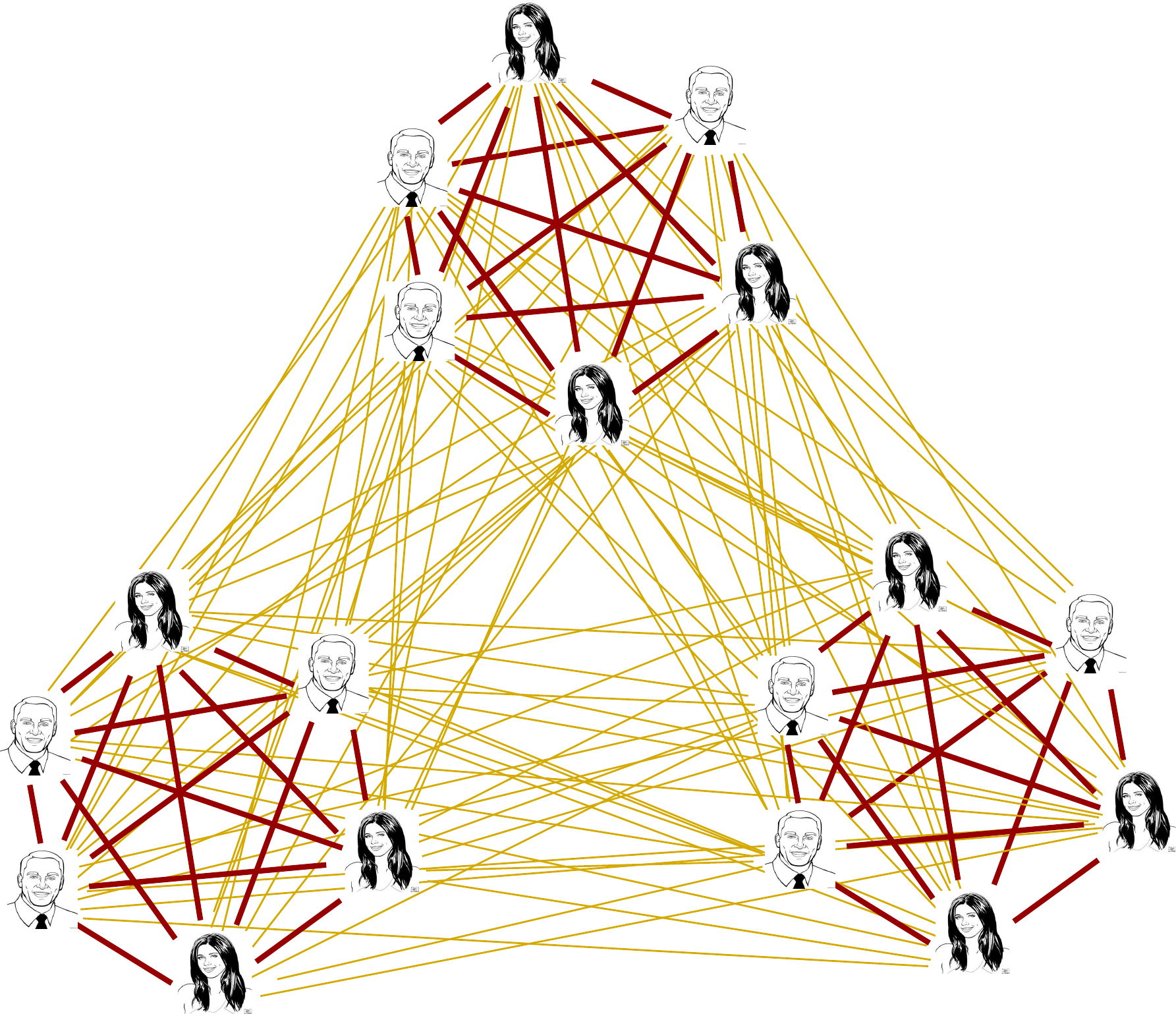}%
\\%
\parbox[t]{0.48\textwidth}{%
\centerline{(a)}%
}%
\hfill%
\parbox[t]{0.48\textwidth}{%
\centerline{(b)}%
}%
 \caption{(Color online) Example of an ego-centric social network (a) with two layers of friends, and an example of the corresponding entire hierarchical social network of a minimal size (b). Dark links (red online) connect the first layer friends and light-grey links (yellow online) correspond to the second layer friends.} \label{fig_h1}
\end{figure}

We generate several networks of social interactions that satisfy the following conditions:
i) each individual has a number of ego-centric friendship layers up to the third, fourth or fifth levels; and
ii) the sizes of these inclusive layers increase with the scaling ratio 3, thus consisting of 5, 15, 45, 135 etc. members.
An example of such an ego-centric network is shown in figure~\ref{fig_h1}~(a).
Since the entire networks with the given ego-centric picture may be designed in a number of ways, we generate two types of networks that satisfy described the ego-centric properties.

Initially, the corresponding random networks were generated. For this reason, we generate a set of nodes the number of which is given in table~\ref{tab1}.
\begin{table}[!ht]
\caption{Basic network characteristics: the number of nodes $N$ and the total number of links $L$ for the 3-, \mbox{4-,} and 5-layer networks. The given number of nodes $N$ is the minimal number of nodes that are necessary to reproduce the network with the described ego-centric picture.} \vspace{2ex}\label{tab1}
\begin{center}
\tabcolsep=0.5em
\begin{tabular}{|c|r|r|}
\hline
            & $N$ &   $L$  \\\hline\hline
  3 layers  &  54 &   1251 \\\hline
  4 layers  & 162 &  10935 \\\hline
  5 layers  & 486 &  98415 \\\hline
\end{tabular}
\end{center}
\end{table}
Then, each individual $i$ is connected to randomly chosen five other nodes in the system so that each ego arrives at 5 first layer neighbors connected by the strongest links of weight $\alpha_1$. Then, we repeat this procedure for each consecutive layer $l$ (link weight $\alpha_l$) avoiding having multiple links between any pair of nodes. Thus, each an individual is expected to have 10 links of weight $\alpha_2$ to the second layer friends, 30 links of weight $\alpha_3$ to the third layer friends, etc. The expected number of links $L$ in a network is shown in table~\ref{tab1}. Due to this construction procedure, a few nodes may have a smaller number of the weakest links in the network. However, we expect that this does not affect the final result since this number is small enough and concerns only the weakest links in a random network.

Then, as another alternative for a network with ego-centric property, we construct a hierarchical network. Initially, the entire set of nodes (see table~\ref{tab1}) is divided into groups of six individuals each, and each pair of the nodes within each group are connected by the links of the strongest weight $\alpha_1$. Thus, each ego has got 5 links to the first layer friends. Next, the set of these groups are grouped into clusters, each consisting of 3 groups and thus containing altogether 18 nodes. In order to build the second layer of ego-centric network, each node of the initial group is connected to 5 out of 6 individuals of each of the selected groups in the cluster with the link weight $\alpha_2$. The choice of the five nodes is done in the way to obtain a maximally regular network. The example of such network is shown in figure~\ref{fig_h1}~(b). In order to obtain a hierarchical network of higher order, we repeat the procedure by merging the three clusters, etc. The constructed hierarchical network has a strong relation between the number of nodes $N$ and the number of links $L$ in the network, see table~\ref{tab1}.

The distribution of link weights between an ego and the members of each layer $l$ of his or her ego-centric network shows a proportional decrease of the link weight $\alpha_l$ with the layer number $l$ with a decreasing rate:
\begin{equation}\label{eq_rate}
r=\alpha_l/\alpha_{l+1}\,.
\end{equation}
Having generated these two different networks of social interactions, we will next perform numerical simulations of the  biased transmission model on them with different values of $r$.

\subsection{Biased transmission model}
Now let us consider the model of biased transmission equation~(\ref{eq2.4}) on the above-discussed network structures.
In the simulations, we use the same value $\alpha_0=0.5$ for each individual $i=1,\ldots, N$ in the system, which gives the probability that an individual $i$ considers him- or herself as a source of cultural trait in unbiased case $b=0$. The bias parameter is set fixed to $b = 0.1$ and we use three different values of parameter $r$ (\ref{eq_rate}): $r=\{2, 5, 10\}$. For the initial ($t=0$) conditions, only a single randomly chosen individual $k$ is characterized by a favored cultural trait $x_k(0) = 1$ and all the others $m\neq k$ are characterized by $x_m(0) = 0$.
The results of numerical simulations are averaged over 1000 runs of the model and a random network has been regenerated for each run.
Since initially only a single node $k$ is characterized by a more attractive cultural trait $x_k(0)=1$, the favored trait has disappeared in more than $60\%$ of runs at the
\begin{figure}[!h]
\begin{center}
\includegraphics[width=100mm]{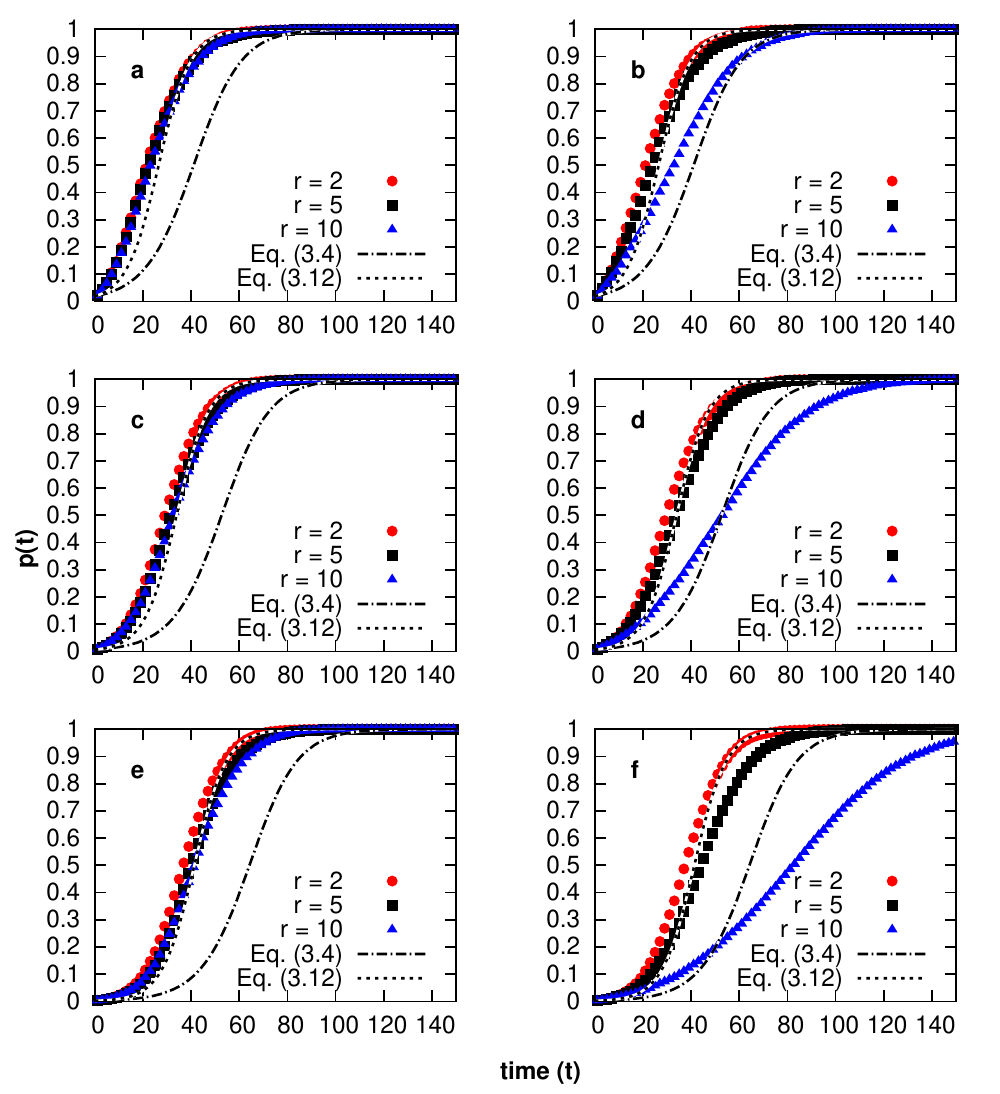}
\end{center}
 \caption{(Color online) Fraction $p(t)$ of trait favored by bias as a function of time $t$. Top panels [(a), (b)] correspond to three layer ego-centric network of social interactions, panels (c) and (d) correspond to the four layer network, and the bottom panels (e) and (f) to the five layer networks. The left hand side panels demonstrate evolution of cultural trait in random networks while the right hand side panels correspond to the hierarchical networks. The evolution of cultural trait on networks with different values of link weight rate $r=\{2, 5, 10\}$ are represented by balls, squares and triangles, respectively. Dot-dashed line corresponds to the results of the single external source mean-field approximation (\ref{eq3.4}) and the dotted line represents the results of the mean field approximation (\ref{eq014}) for layered ego-centric networks.
 } \label{fig_2}
\end{figure}
early stage of simulations. Thus, below we consider only those runs at time $t$ for which the system contains at least a single node with the trait $1$ at $t$. The results of our investigation are shown below for networks with the different number of layers.

Let us start with 3-layer egocentric network, where each individual has 5 first-, 10 second- and 30 third-layer friends, so that inclusive layer sizes follow 5, 15 and 45. The fraction $p(t)$ of individuals with a favored cultural trait on a random network is shown in figure~\ref{fig_2}~(a) as a function of time $t$.
The same dependency on a hierarchical network with 3-layer ego-centric networks is shown in figure~\ref{fig_2}~(b). The mean field approximations (\ref{eq3.4}) and (\ref{eq014}) are shown by dot-dashed and dotted lines, respectively. The results show that the approximation (\ref{eq014}) describes the evolution of a cultural trait for a random network much better than the single source approximation \cite{Boyd1985}. Moreover, the distribution of the link weights among the layers, described by parameter $r$, does not affect the evolution of a cultural trait in random networks as predicted by the mean-field approximation. However, the value of $r$ affects the process on hierarchical networks, so that for large values of $r$, the evolution of the favored cultural trait in the hierarchical network becomes slower than expected by mean field approximation.

This behavior is even more pronounced for the systems having four and five layers in ego-centric social networks. The results for the four-layer ego-centric networks, where each individual has 5 first-, 10 second-, 30 third- and 90 fourth layer friends are depicted in figure~\ref{fig_2}~(c) and (d). Note that in this case the last inclusive layer contains 135 external sources corresponding  to a rough upper limit of active human relations an individual can maintain.

For the sake of completeness, we also consider network structures that contain five layers where the last layer contains 270 friends. Here, the inclusive fifth layer with the total of 405 friends goes beyond the Dunbar number limit, thus including individuals who are not actively related to the central ego. The results of the spreading simulations for the hierarchical and random networks are shown in figure~\ref{fig_2}~(e) and \ref{fig_2}~(f) for comparison. These results confirm the previously described behavior of the 3-layer ego-centric networks, so that the slowing down effect is even further enhanced in hierarchical networks for large values of $r$.

\section{Conclusions}\label{sect_V}

To summarise, we have here developed a model of transmission of cultural traits in layered and weighted egocentric networks. The network structure was generated either randomly or to have regular hierarchical structure in accordance with the ego-centric friendship layering showing decreasing emotional closeness with an increasing distance from the ego, as suggested by Robin~Dunbar. We studied this model first analytically with two types of mean field approximations, namely one based on a single external source of cultural trait and the other on a layered ego-centric network. In order to validate these mean field results, we compared them with the results obtained by extensive numerical simulations, and found them behaviorally similar, but especially the approximation of the layered ego-centric network seemed to compare well with the simulation results.

In the simulation studies, we observed a distinct slowing down effect in the evolutionary process of a cultural trait. This may sound somewhat surprising against the fact that social networks show a small world topology \cite{Barabasi2003}, reflecting that any individual is only a few steps away from any other. However, a similar behavior is seen in a spreading process following susceptible - infected dynamics, which, on the one hand, was caused by the inner community structure of the social network and, on the other hand, by the bursty nature of social interactions between a pair of individuals \cite{Karsai2011}. Similar mechanisms were found responsible in maintaining long-term cultural diversity beside the emergent short-term collective behavior \cite{Valori2011}. The origins of these kinds of systemic behavior bear close resemblance with the critical slowing down round the critical phase transition points of materials where regardless of the long range correlations, the existence of ordered domains slow down the relaxation of the whole system towards the stable state.

The slowing down in our study is related to the appearance of a well defined community structure or network clusters. The clustered structure for random networks does not exist by definition and the obtained mean-field approximation is capable of properly describing the evolution of the cultural trait in a social system. However, the appearance of a clustered structure in hierarchical networks for large values of the control parameter $r$ results in slowing down that cannot be captured by the mean-field approximation. On the one hand, this limits the validity of the derived approximation (\ref{eq014}) and, on the other hand, underlines the importance of community structure for the spreading and evolution processes in social systems. Besides the importance of the structure of individual ego-centric social networks in creating channels to transmit cultural trials, the composition of these ego-centric networks together with the whole system play an important role in the functionality of the system.

\section*{Acknowledgements}
This paper is dedicated to the memory of A.~Olemskoi, a physicist whose research was closely related to synergy and investigation of complex systems. VP acknowledges the good memory of writing the first review paper \cite{Holovatch2006} on complex networks in Ukrainian together with A.~Olemskoi.

This work was supported by the FiDiPro program  (TEKES, Finland). JK acknowledges support from FuturICT.hu (grant no.: T\'AMOP--4.2.2.C--11/1/KONV--2012--0013).

\newpage
\ukrainianpart

\title{Перенесення культурних особливостей у пластоподiбних егоцентричних мережах}
\author{В. Пальчиков\refaddr{label1,label2,label3}, К. Каскi\refaddr{label1,label4,label5}, Я. Кертес\refaddr{label6,label1}}
\addresses{
\addr{label1} Вiддiл бiомедичних технологiй та комп’ютерних наук, Унiверситет Аалто, Фiнляндiя
\addr{label2} Iнститут фiзики конденсованих систем НАН України, вул. Свєнцiцького, 1, Львiв, Україна
\addr{label3} Iнститут теоретичної фiзики iм. Лоренца, Лейденський унiверситет, Лейден, Нiдерланди
\addr{label4} CABDyN центр, Бiзнес школа Саїда, Оксфордський унiверситет, Оксфорд, Великобританiя
\addr{label5} Центр дослiдження складних систем та вiддiл фiзики, Пiвнiчно-схiдний унiверситет, Бостон, США
\addr{label6} Центр наук про мережi, Центрально-європейський унiверситет, Будапешт, Угорщина
}
\makeukrtitle

\begin{abstract}
\tolerance=3000%
Хоча було розвинено ряд моделей для дослiдження появи культури та еволюцiйних фаз у соцiальних системах, один важливий аспект не був достатньо висвiтлений. Цим аспектом є роль структури мереж соцiальних зв’язкiв, якi служать каналами для передачi культурних особливостей i якi вiдiграють важливу роль в еволюцiйних процесах у соцiальних системах. У данiй статтi ми дослiджуємо цю роль у моделi переачi культурних особливостей та їх еволюцiю в соцiальних системах, спираючись на пластоподiбнi структури егоцентричних мереж, надхненi гiпотезою соцiального мозку. Для цiєї моделi ми отримали аналiтичний розв’язок у дусi середнього поля та дослiдили його застосовнiсть, порiвнюючи отриманi передбачення з результатами чисельних симуляцiй.
\keywords егоцентричнi мережi, еволюцiя культури, наближення середнього поля
\end{abstract}

\end{document}